\documentclass[]{interact}

\usepackage{epstopdf}
\usepackage{subfigure}
\usepackage{amsmath}
\usepackage{amsfonts}
\usepackage{amssymb}
\usepackage{mathtools}
\usepackage{todonotes}
\usepackage{soul} 
\usepackage{hyperref}

\usepackage[numbers,sort&compress,merge]{natbib}
\bibpunct[, ]{(}{)}{,}{n}{,}{,}

\theoremstyle{plain}

\theoremstyle{definition}

\theoremstyle{remark}

\begin{document}
	
\articletype{ARTICLE}

\title{Four-wave mixing in a non-degenerate four-level diamond configuration in the hyperfine Paschen--Back regime}

\author{
\name{Daniel J. Whiting, Renju S. Mathew\thanks{CONTACT Renju S. Mathew Email: r.s.mathew@durham.ac.uk}, James Keaveney, Charles S. Adams, Ifan~G.~Hughes}
\affil{Joint Quantum Centre (JQC) Durham-Newcastle, Durham University, Department of Physics, South Road, Durham, DH1 3LE, United Kingdom}
}

\maketitle

\begin{abstract}
We present an experimental study of seeded four-wave mixing (4WM) using a diamond excitation scheme (with states from the 5S$_{1/2}$, 5P$_{1/2}$, 5P$_{3/2}$ and 5D$_{3/2}$ terms) in a thermal vapour of $^{87}$Rb atoms. We investigate the 4WM spectra under the application of a strong magnetic
field (0.6 T). The Zeeman interaction is strong enough to realise the hyperfine Paschen-Back regime, which has the effect of separating the optical transitions by more than the Doppler width, thereby significantly simplifying the spectral features. We show that this facilitates a quantitative comparison, even in the regime of strong dressing, between experimental data and a simple theoretical model based only on four-level optical Bloch equations.

\end{abstract}


\section{Introduction}

The non-linear optical process of four-wave mixing (4WM) continues to generate much interest within the atomic physics community, with a growing list of applications that include: producing correlated photon-pairs~\cite{Willis2011} for use in quantum information protocols; relative intensity squeezing~\cite{Mccormick2007}; creating entangled imaging systems~\cite{Boyer2008}; creating collective spin excitations~\cite{MacRae2012}; observing collective quantum beats~\cite{Whiting2017} and transferring trans-spectral orbital angular momentum~\cite{Walker2012}. Seeded 4WM has also been widely utilised for many applications, including: precision spectroscopic measurements~\cite{Ribeiro1997}; displacement measurements in electro-mechanical cantilevers~\cite{Pooser2015}; investigations in pulse-seeded Rydberg systems of motional dephasing~\cite{Chen2010} and photon storage~\cite{Ripka2016}.

Atomic vapours are a natural fit for many of these applications, particularly in quantum information, since the photons that are produced are inherently frequency- and bandwidth-matched to other elements of the system, e.g. atom-based quantum memories~\cite{Heshami2015}; quantum repeaters~\cite{Sangouard2011} or quantum gates~\cite{Paredes-Barato2014a}.
%
Various arrangements of energy levels have been used to generate 4WM: the double lambda~\cite{Hemmer1995a,Nice2003,Kuzmich2003,Polyakov2004,MacRae2012}; double ladder~\cite{Parniak2016,Leszczynski2016,Lee2016c} and the diamond~\cite{Whiting2017,Becerra2008,Willis2009a,Parniak2015} schemes being the most widely reported.

However, 4WM is complicated by multi-level effects due to hyperfine structure, particularly in thermal atomic vapours where the Doppler width exceeds the atomic hyperfine splittings. This makes modelling of these systems challenging and computationally demanding, and can obscure the underlying physics.  To date, models which give good agreement between theory and experiment require a restriction on the strength of the driving fields and a compromise between simplicity and accuracy \cite{Parniak2015}. 

The goal of our work is to realise a 4WM signal in a nondegenerate system, thereby allowing for accurate quantitative modelling over a wider range of experimental parameters.
One possible method is to use optical pumping to reduce the number of participating atomic energy levels, but this significantly adds to the experimental complexity.  It is also  
 practically challenging to realise optical pumping  in thermal vapours, and typically requires further complexities such as the use of buffer gases and/or anti-relaxation coatings~\cite{Robinson1982,Seltzer2009a} to reduce the effect of state-changing collisions.
%
An alternative approach, which we demonstrate in this work, is to apply a large magnetic field. At fields where the Zeeman shift exceeds the internal hyperfine energy intervals of the atom, known as the hyperfine Paschen-Back (HPB) regime, the degeneracy of the transitions is lifted~\cite{Olsen2011,Weller2012a,Weller2012,Sargsyan2014,Zentile2014a,Sargsyan2015b,Sargsyan2017, Sargsyan2017a}.

In a weak external magnetic field, the nuclear spin $\mathcal{I}$ and the electron angular momentum $\mathcal{J}$ couple to give the total angular momentum $\mathcal{F}$; the latter, and its projection on to the field axis, $m_{F}$, are good quantum numbers.  Decoupling of  $\mathcal{I}$ and  $\mathcal{J}$ occurs for stronger fields, and these vectors precess independently about the magnetic field; their projections $m_{I}$ and $m_{J}$ are the good quantum numbers in this regime.  An estimate for the decoupling field, $B_0$, above which the hyperfine Paschen-Back regime is achieved is given by $B_0=A_{\rm hfs}/\mu_{\rm B}$, where $A_{\rm hfs}$  is the ground-state hyperfine coupling coefficient, and $\mu_{\rm B}$ is the Bohr magneton.  For $^{87}$Rb, $B_0=0.25$~T. 
 We utilise a 0.6~T magnetic field to enter the HPB regime and comprehensively characterise four-wave mixing in a diamond scheme in a thermal Rb vapour.  Further, we show that in this regime the observed 4WM signals can be mapped to an isolated four-level system that can be easily modelled using only simple 4-level optical Bloch equations. 

\begin{figure}
	\begin{minipage}{\textwidth}
		\includegraphics[width=\columnwidth]{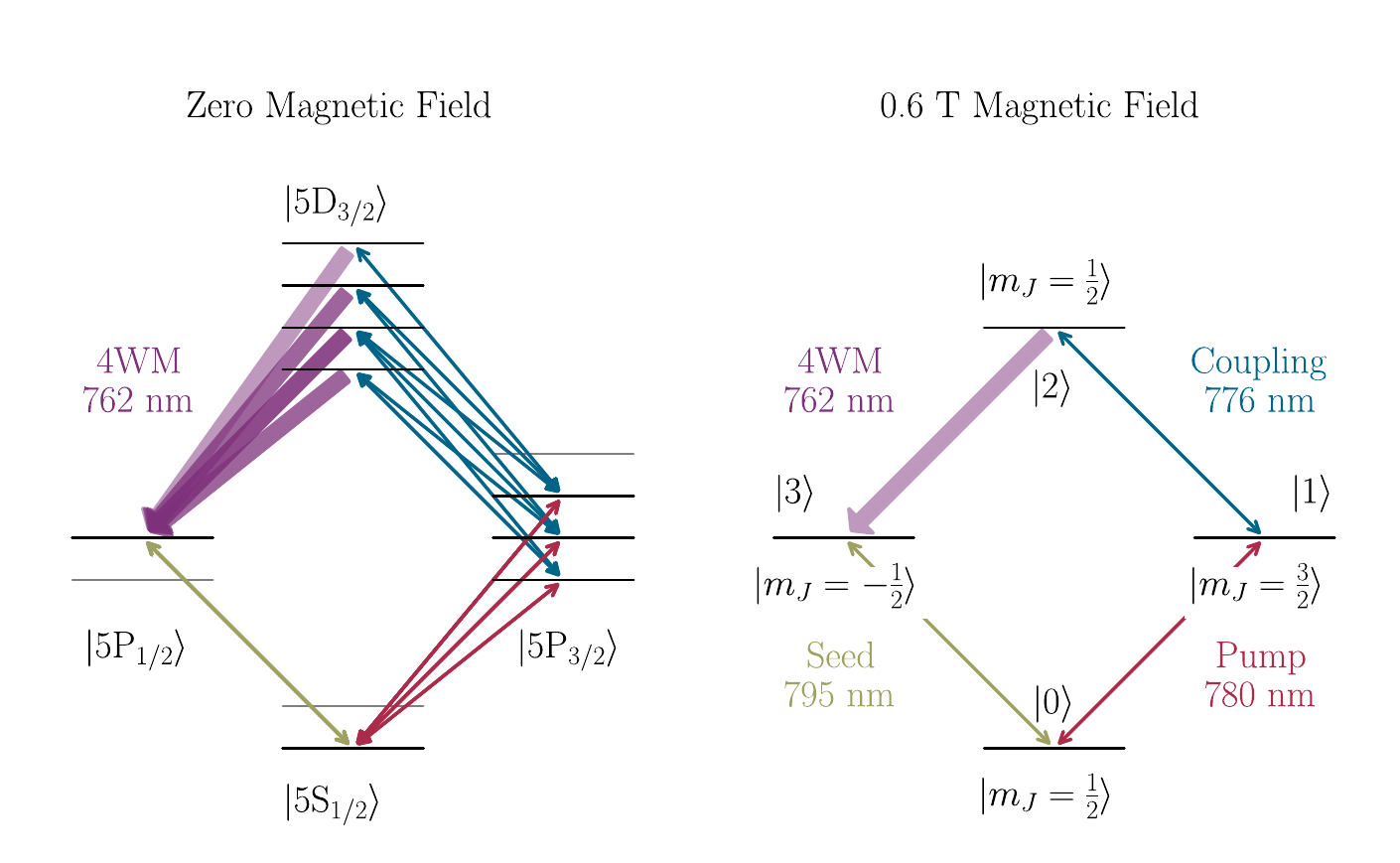}
		\caption{Rubidium energy levels in the diamond-scheme in the absence (left) and presence (right) of a magnetic field of strength 0.6 T.
			Three beams (pump, seed \& coupling) are added. When the phase matching condition is fulfilled, the four-wave mixing process generates a fourth beam (4WM). In the absence of a magnetic field, there is interference along multiple paths; the magnetic field removes the multiple-path interference. The states labelled 0--3 are those used in the model.} \label{energy-levels}
	\end{minipage}
	\vspace{.75em}
	\begin{minipage}{\textwidth}
	\includegraphics[width=\columnwidth]{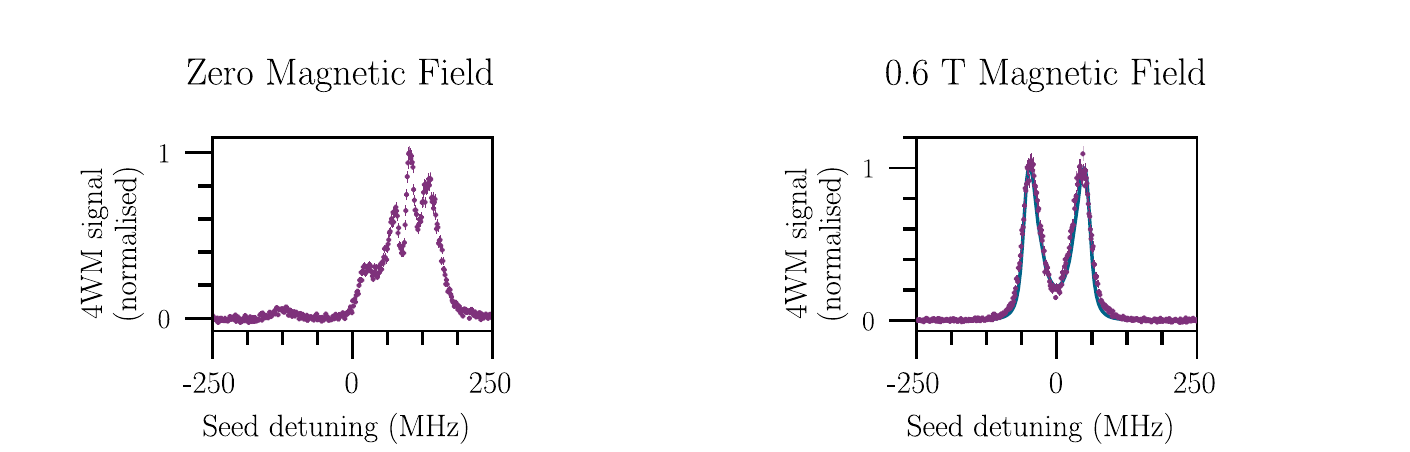}
	\caption{Example experimental four-wave mixing spectra in the diamond-scheme in the absence (left) and presence (right) of a magnetic field of strength 0.6 T. In its absence, the four-wave mixing spectra are both highly sensitive to experimental conditions and difficult to model. Applying a magnetic field results in textbook four-wave mixing spectra that can be quantitatively modelled.} \label{example-spectra}
\end{minipage}
\end{figure}

In Fig. \ref{energy-levels} we show the general principle of the experiment. In the absence of an applied magnetic field (left), the hyperfine splitting of the 5S$_{1/2}$, 5P$_{3/2}$, 5P$_{1/2}$ and 5D$_{3/2}$ terms creates many paths for generating a 4WM signal. Four-wave mixing is a coherent process where the electric fields from different paths interfere with one another, complicating the observed 4WM signal which becomes difficult to predict. In a thermal atomic vapour, the hyperfine splitting of the 5P and 5D states is smaller than the Doppler width, so many sub-states are simultaneously excited amongst different atomic velocity groups. Additionally, although strong pumping (with high Rabi frequencies) is often desirable in order to obtain high conversion efficiencies, Rabi-frequencies similar to the hyperfine structure splitting can also lead to this multi-path interference~\cite{Parniak2015, Badger2001}. The left panel of Fig. \ref{example-spectra} shows an example 4WM spectrum illustrating the effect of multiple excitation pathways.
However, when a large enough magnetic field is applied (the HPB regime), adjacent levels in the 5S manifold are separated by much more than the Doppler width and, due to selection rules for electronic transitions, individual two- \cite{Weller2012,Zentile2014a}, three- \cite{Whiting2015,Whiting2016a} and four-level systems can be coupled separately, as shown on the right of Fig. \ref{example-spectra}. The 4WM signal in this case is much simpler, consisting of an Autler-Townes split doublet that results from the strong dressing by the coupling laser.

\section{Experimental details}

\begin{figure}[t]
	\centering
	\includegraphics[width=\columnwidth]{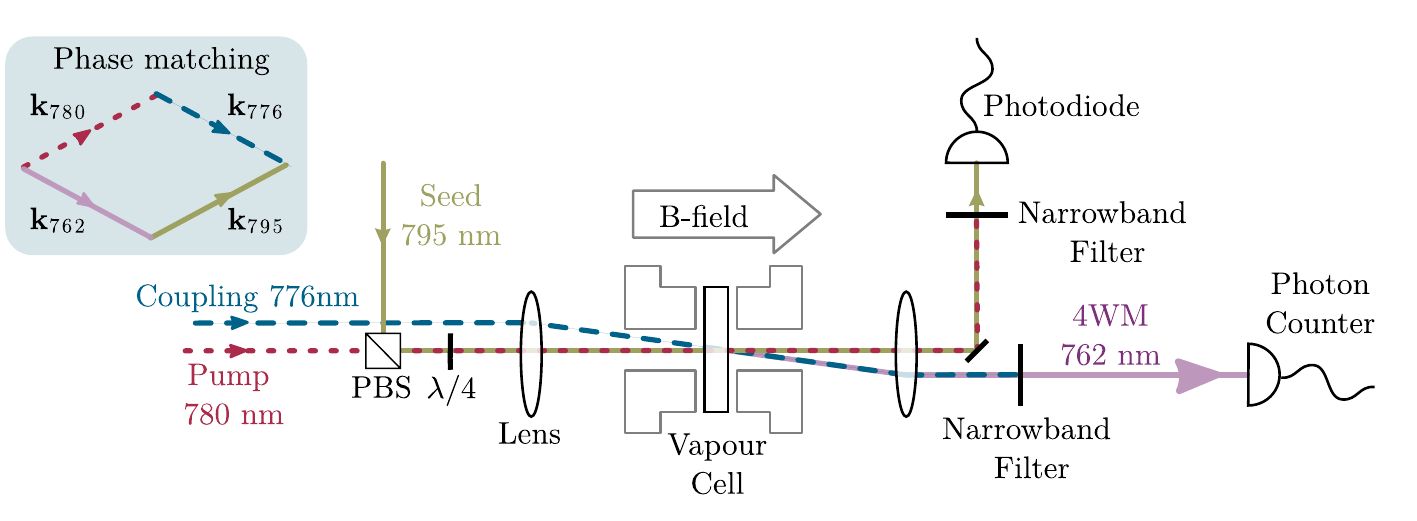}
	\caption{
		Schematic of the experimental configuration. Three beams are focused through a heated vapour cell of length 2 mm containing isotopically enriched $^{87}$Rb in a uniform magnetic field of strength B = 0.6~T along the pump beam axis. The coupling beam crosses the pump beam at an angle of 10 mrad  (not to scale). The phase-matching condition is fulfilled as shown in the inset. The pump and seed polarisations are set by a quarter-waveplate ($\lambda/4$). The seed transmission is measured using a photodiode and the generated four-wave mixing (4WM) field passes through a narrowband interference filter and is detected on a photon counter. 
	}
	\label{exp-setup}
\end{figure}

A schematic of the experimental setup is shown in Fig. \ref{exp-setup}. We use a 2~mm long vapour cell containing isotopically enriched rubidium ($>98\%~^{87}$Rb), which is placed between two cylindrical NdFeB magnets --- the figure shows a cross-sectional view of the top-hat-profile of the magnets. We achieve a magnetic field across the vapour cell of 0.6~T, which is uniform to the 1~mT level (the field profile is shown in Fig. 1 of reference \cite{Whiting2015}). The vapour pressure, and hence the atomic number density, is controlled via the cell temperature. In the magnetic field, the circularly polarised pump beam at 780~nm drives the $\sigma^+$ transition between the  $|$5S$_{1/2}$, $m_J=\frac{1}{2}\rangle$ and $|$5P$_{3/2}$, $m_J = \frac{3}{2}\rangle$ states. The pump beam is overlapped in the cell with the coupling (776~nm) and seed (795~nm) beams, which are resonant with the $|$5P$_{3/2}$, $m_J = \frac{3}{2}\rangle \rightarrow |$5D$_{3/2}$, $m_J = \frac{1}{2}\rangle$ and \mbox{$|$5S$_{1/2}$, $m_J=\frac{1}{2}\rangle \rightarrow |$5P$_{1/2}$, $m_J = -\frac{1}{2}\rangle$} transitions, respectively. $m_I = \frac{3}{2}$ in all cases.
All beams are focussed to achieve the necessary high intensities with a $1/e^2$ radius of $\sim60$~$\mu$m at the centre of the cell. The phase-matching condition is fulfilled as shown in the inset of Fig. \ref{exp-setup}. The pump and seed beams are co-propagating, whilst the coupling beam crosses at a small angle so that the signal and seed beams can be easily separated from the pump and coupling beams using narrowband interference filters. The signal beam is detected using a photon counting module and the transmitted seed light is measured on a conventional photodiode. The pump and seed powers are 1 $\mu$W whilst the coupling power is 34 mW.

Changing the alignment of the beams modifies the lineshapes; we conjecture that this is due to the presence of back-reflections since the cell is not anti-reflection coated. We optimise the alignment to ensure that the Autler-Townes splitting is maximised.

\section{Four-level model}
\vspace{-.75em}

\begin{figure}[t]
	\centering
	\includegraphics[width=0.95\columnwidth]{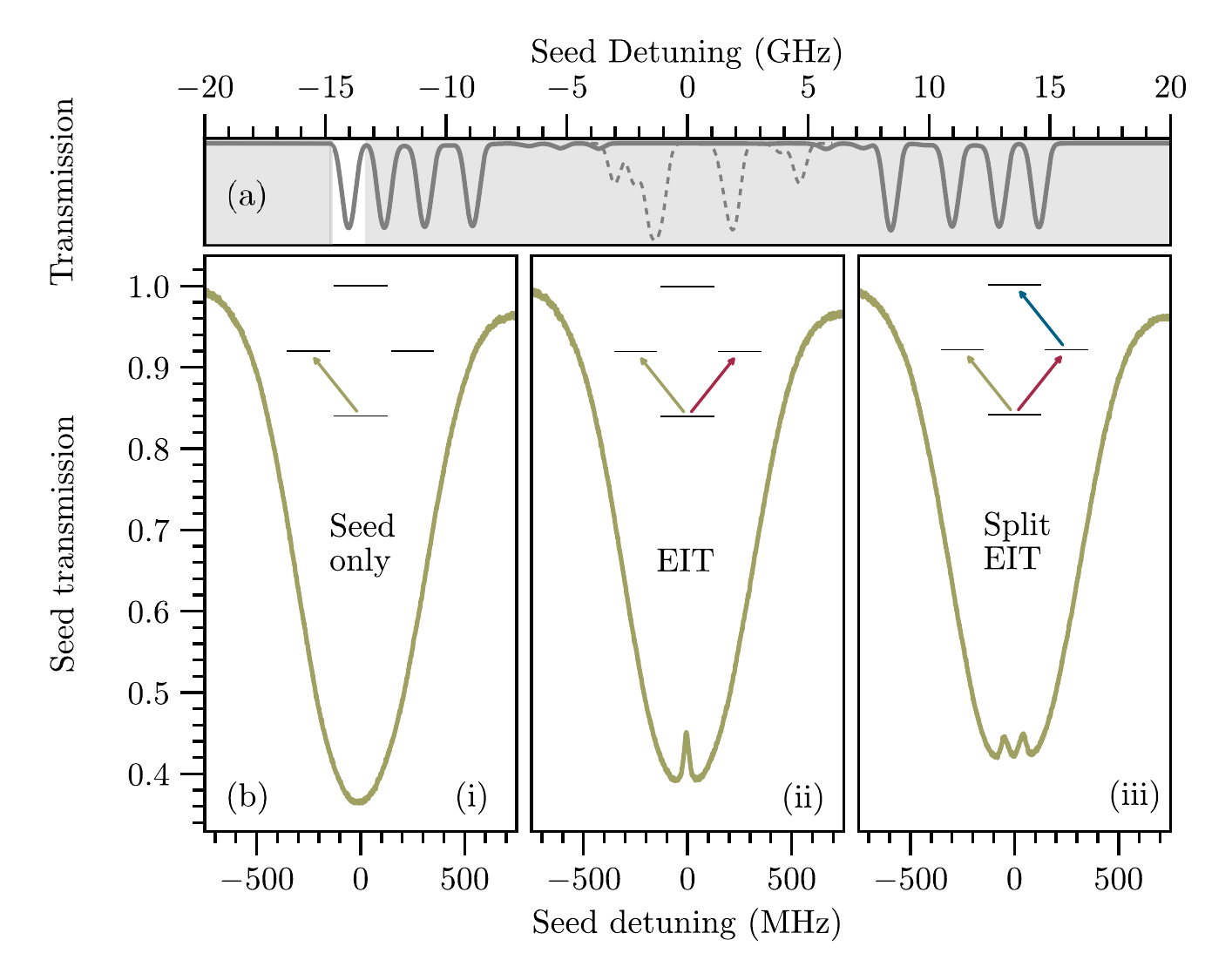}
	\caption{
	(a) Theoretical transmission spectrum of the rubidium D$_1$ (795 nm) line in the presence (thick grey line) and absence (dotted grey line)  of a 0.6 T magnetic field at a temperature of 80$^{\circ}$C. The unshaded region highlights the detuning range for the experimental data shown below. (b) Experimental transmission spectra of the 795 nm seed light in (i) the presence of the seed beam only (ii) the seed and pump beams and (iii) the seed, pump and coupling beams. (b)(iii) highlights the role of the dressed states in this configuration.} 
	\label{EIT-split-EIT}
\end{figure}

We model a system of four-levels $|0-3\rangle$, labelled anti-clockwise from the ground-state in Fig. \ref{energy-levels}, interacting with three CW driving fields with Rabi frequencies $\Omega_{780}$, $\Omega_{776}$, $\Omega_{795}$ and detunings $\Delta_{780}$, $\Delta_{776}$, $\Delta_{795}$. Note that these are angular detunings but in the figures of this paper we use linear detunings i.e. $\Delta / 2\pi$.  We write the interaction Hamiltonian (in the rotating wave approximation) as
\begin{equation*}
\hat{H} = \frac{\hbar}{2}
\begin{pmatrix}
0 & \Omega_{780} & 0 & \Omega_{795} \\
\Omega_{780} & -2\Delta_{780} & \Omega_{776} & 0 \\
0 & \Omega_{776} & -2(\Delta_{780}+\Delta_{776}) & 0 \\
\Omega_{795} & 0 & 0 & -2\Delta_{795} \\
\end{pmatrix}
\end{equation*}
and solve the Lindblad master equation to find the steady-state density matrix $\rho$. For the sake of simplicity we do not explicitly write the phase of the driving fields in $\hat{H}$. In related experiments where the medium is driven by 4 driving fields, the phases need to be considered carefully as they lead to complex interference effects \cite{Morigi2002}. However in our experiment the relative phase of the 3 driving fields only determines the phase of the generated 4WM field, which is not measured.
The radiated electric field on the $|2\rangle\rightarrow|3\rangle$ transition is proportional to the off-diagonal matrix element $\rho_{23}$. The atomic motion is accounted for by considering the Doppler shifted detunings for each velocity class $v$. Due to phase-matching the total radiated electric field is the coherent sum over all $\rho_{23}(v)$ weighted by the Maxwell-Boltzmann velocity distribution. In order to accurately model the experimental spectra we should also consider the spatial intensity profiles of the driving fields. We do this by assuming that all of the driving fields have Gaussian intensity profiles in the radial direction $r$ (which is consistent with our measurements) and are perfectly overlapped. The radiated field is then calculated for radial shells over which constant intensities can be assumed. In the experiment we detect the signal using a single-mode optical fibre and a photon counting module. The effect of the optical fibre collection is modelled as a Gaussian spatial filter acting on the emitted 4WM field and therefore we calculate the theoretical signal amplitude as
\begin{equation}
S \propto \int_{0}^{\infty} |\rho_{23}|^2 \exp(-2r^2/w_{\rm D}^2)~r\mathrm{d}r
\label{signal}
\end{equation}
where $w_{\rm D} \sim90$ $\mu$m is the measured waist of the optical fibre detection mode at the position of the cell.

\begin{figure}[t]
	\centering
	\includegraphics[width=0.9\columnwidth]{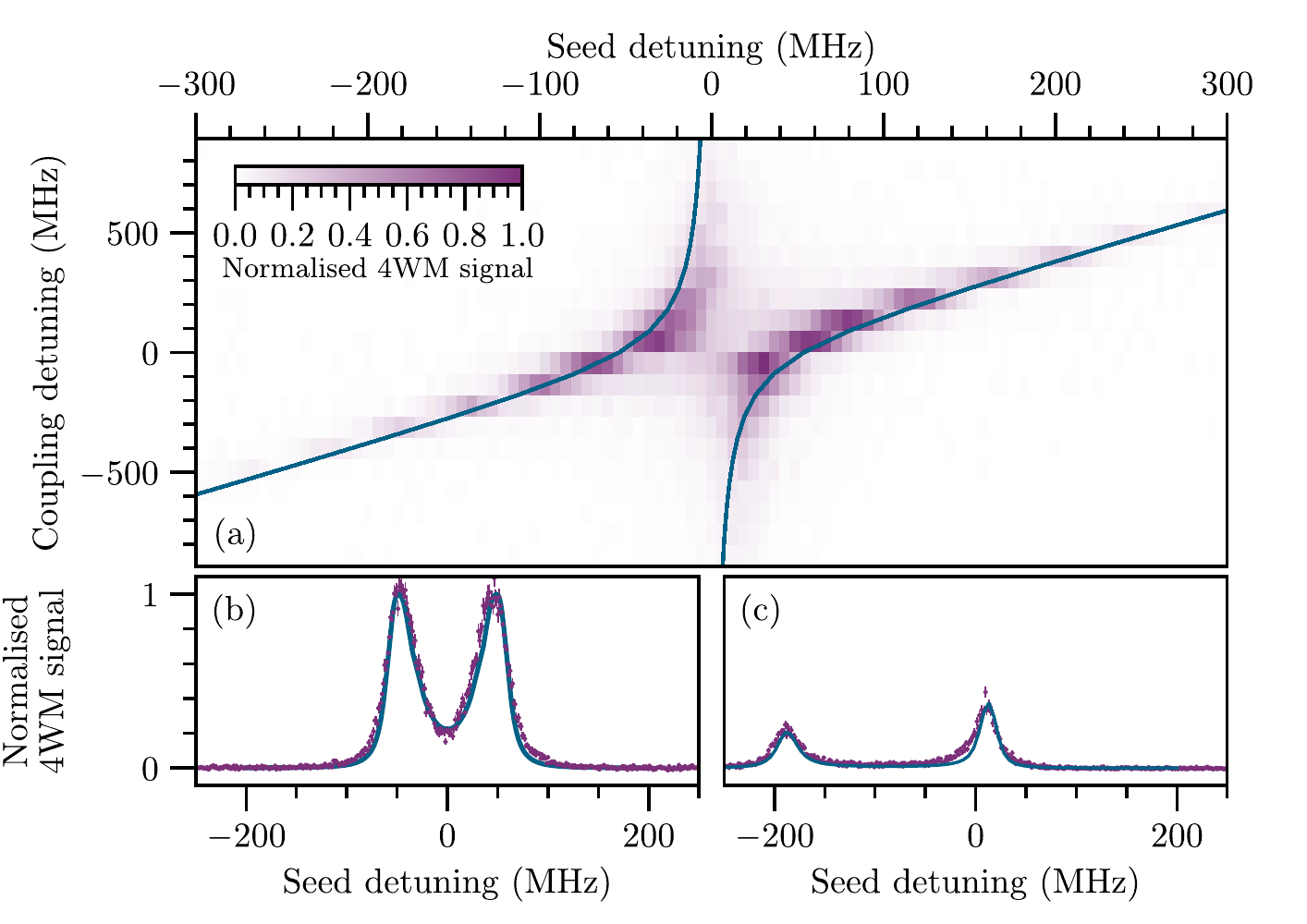}
	\caption{Four-wave mixing signal as a function of both coupling (776 nm) and seed beam (795 nm) detunings. The system exhibits an avoided crossing, which is characteristic of Autler-Townes splitting. The solid lines in the panel (a) are a plot of $\Delta_{795} = k_{795}v_{\rm 2 photon}$ where $v_{\rm 2 photon}$ is given by Eq. (\ref{v2photon}). Panels (b) and (c) show the 4WM signals at coupling detunings of 0 MHz and -360 MHz respectively, where the solid lines are the model as given by Eq. (\ref{signal}) with $\alpha = 35$ and $\Omega_{776}$ reduced by 14\% from the measured value.} 
	\label{4WM-vs-detuning-coupling-and-seed}
\end{figure}
\section{Results}

Fig. \ref{EIT-split-EIT} shows a typical spectrum as the seed laser frequency is scanned. In panel (a) we show, for reference, a theoretical transmission spectrum \cite{Zentile2015b, keaveney2017elecsus} of the Rb D$_1$ line with (solid line) and without (dashed line) the applied magnetic field. Note that the spectrum without field is for a naturally abundant vapour cell while the spectrum with field is for an isotopically enriched $^{87}$Rb cell. Zero detuning is the weighted D$_1$ line centre of naturally abundant rubidium in zero magnetic field \cite{Siddons2008b}. In panel (b)(i)
we scan the seed laser over the leftmost absorption line (unshaded region in panel (a)) and measure its transmission spectrum.  Turning on the resonant 780~nm pump beam (b)(ii) creates a V-system which exhibits a small peak at line-centre due to electromagnetically induced transparency (EIT). When the strong coupling laser is also turned on (b)(iii), it dresses the pump transition and the EIT feature is split into an Autler-Townes doublet, highlighting the role of the dressed states in this configuration.

Having monitored the transmission of the seed (with the photodiode in Fig. \ref{exp-setup}) to characterise the system, we now move on to measuring of the 4WM signal detected with the photon counter.
The combined use of narrowband interference filters and photon-counting modules to detect the 4WM signal yields a high signal-to-noise ratio, as is evident in Fig. \ref{example-spectra} (right). This is necessary because we use low temperatures and low pump and seed powers, resulting in a small absolute 4WM signal (pW).

We now look at how the observed 4WM signal changes with a selection of the experimental parameters. The complete parameter space is too large to fully explore here, but we show the results of a range of experiments along with the theoretical model demonstrating the success of the 4-level model in the HPB regime. 

\subsection{Model parameters}
The decay rates of the relevant atomic states are $\Gamma_{10} = 2\pi\times6$~MHz, $\Gamma_{30} = 2\pi\times5.7$~MHz, $\Gamma_{21} = 2\pi\times0.17$~MHz, $\Gamma_{23} = 2\pi\times0.43$~MHz and $\Gamma_{20} = 2\pi\times0.07$~MHz~\cite{Heavens1961}. In the model, the decay rates $\Gamma_{21}$, $\Gamma_{23}$ and $\Gamma_{20}$ are multiplied by a factor $\alpha$ which includes contributions from collisional buffer-gas broadening on the excited-state transitions~\cite{Sargsyan2010} and the effect of magnetic field inhomogeneity. We find good agreement with the experimental spectra using a value of $\alpha=35$ which corresponds to an additional broadening of $\sim 20$~MHz which is in agreement with previously measured values~\cite{Whiting2016a}. We find it unnecessary to modify $\Gamma_{10}$ and $\Gamma_{30}$, which is unsurprising since the buffer-gas broadening of the 5D states is known to be significantly stronger than of the 5P states~\cite{Sargsyan2010}.

The blue line in Fig. \ref{example-spectra} (right) shows the results of the model using independently determined values of all parameters except for $\Omega_{776}$, which has been reduced by $14\%$, $\alpha=35$ and an overall scaling factor. Given the lack of free-parameters we find that the model is in excellent agreement with the data as indicated by the small residuals. There is clearly some additional broadening which is not currently accounted for by the model. This could be due to imperfect beam overlap or the fact that we do not consider a full propagation model e.g. the Maxwell-Bloch model \cite{Castin1995}.

\subsection{Dependence on coupling beam detuning}
In Fig. \ref{4WM-vs-detuning-coupling-and-seed} we plot the normalised signal level against detuning of both the coupling and seed lasers.

The strong coupling laser dresses the pump transition, creating two pathways to the 5D state which result in the split EIT in  Fig. \ref{EIT-split-EIT} (b)(iii). Hence we observe two features in the 4WM signal that correspond to the resonances with these dressed states.
 For a resonant pump laser ($\Delta_{780} = 0$) and low $\Omega_{780}$, $\Omega_{795}$, the atoms that are two-photon resonant with $|0\rangle\rightarrow|2\rangle$  have axial speeds \cite{Whiting2017}
\begin{equation}
v_{\rm 2 photon} = \frac{\Delta_{776}k_{780}}{a} \pm \sqrt{\frac{\Delta_{776}^{2}k_{780}^{2}}{a^{2}}+\frac{\Omega_{776}^{2}}{2a}}
\label{v2photon}
\end{equation}
where $a = 2(k_{780}^{2}+k_{780}k_{776})$.

In Fig. \ref{4WM-vs-detuning-coupling-and-seed} we plot $\Delta_{795} = k_{795}v_{\rm 2 photon}$ (solid blue line in panel (a)) where $k_{795}$ is the seed transition wavenumber and $\Omega_{776}$ is the measured peak Rabi-frequency reduced by 14\% for the coupling beam. The model agrees reasonably well with the positions of the 4WM resonances, as is expected for seed and pump Rabi-frequencies that are significantly smaller than the coupling Rabi-frequency, but there is a clear overestimation of the Autler-Townes (AT) splitting for $\Delta_{776} = 0 $. This is to be expected since the total 4WM signal is obtained from the entire spatial profile of the coupling field, resulting in a reduced effective Rabi-frequency.

\subsection{Dependence on coupling beam power}
\begin{figure}[t]
	\centering
	\includegraphics[width=0.9\columnwidth]{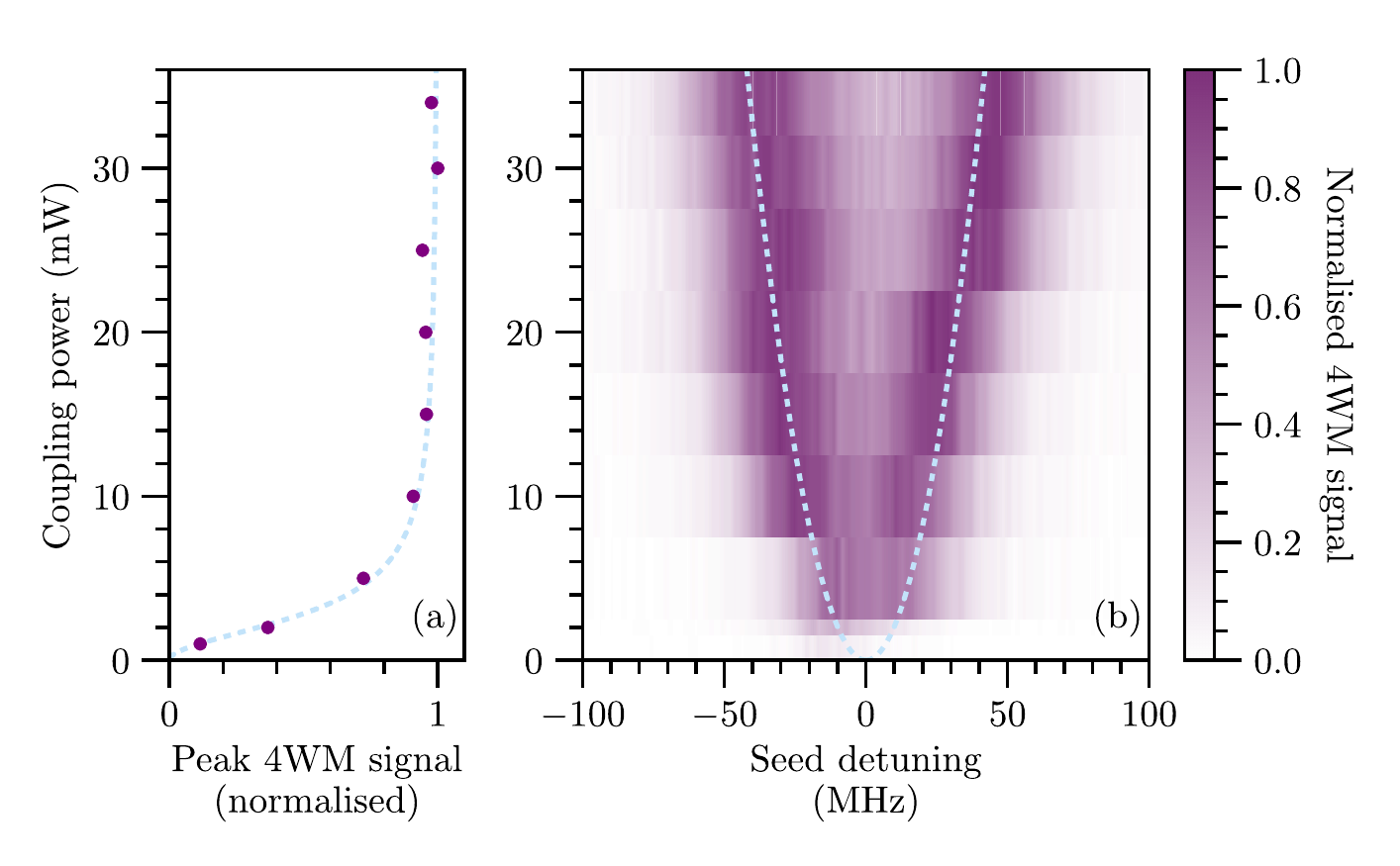}
	\caption{(a) Four-wave mixing signal as a function of the coupling power, with pump and coupling lasers on resonance with the bare atomic transitions. For powers larger than $\sim12$ mW the signal saturates, but the spectral dependence on seed detuning in panel (b) continues to evolve. Autler-Townes splitting is clearly visible as the coupling power increases and shows the expected square root dependence  as shown by the dashed lines. The map has been corrected for laser frequency drift.} 
	\label{4WM-vs-power-coupling}
\end{figure}

The dressed-state resonance frequencies depend on both the coupling detuning (Fig. \ref{4WM-vs-detuning-coupling-and-seed}) and also the coupling laser power, which is plotted in Fig. \ref{4WM-vs-power-coupling}. 
The coupling laser Rabi frequency, $\Omega_C$, is proportional to the square-root of the coupling laser power. 
When the pump and coupling lasers are both on resonance with the bare atomic states, the dressed states are symmetrically split, with the well-known AT energy splitting $\hbar \Omega_C$ \cite{Fleischhauer2005}. 

The dotted lines in Fig. \ref{4WM-vs-power-coupling} show the expected square-root behaviour of the splitting with laser power, which coincide well with the observed peak 4WM signals. Note these dotted lines are a guide to the functional form of the splitting, not a fit to the data.

Note also that since the seed and coupling lasers are not frequency stabilised, there is some drift of the lasers between measurements which we have attempted to correct for in the displayed data.

As the coupling laser power is increased, the dressed-state energy levels split as expected, but the magnitude of the 4WM signal quickly saturates, at a coupling laser power of around 12~mW. This fits well to a simple saturation model given by \mbox{$ 1 - \frac{1}{(1+P^2/A^2)}$} with fit parameter \mbox{$A = 2.8 \pm 0.1$} mW.

\subsection{Dependence on seed beam power}
\begin{figure}[t]
	\centering
	\includegraphics[width=0.8\columnwidth]{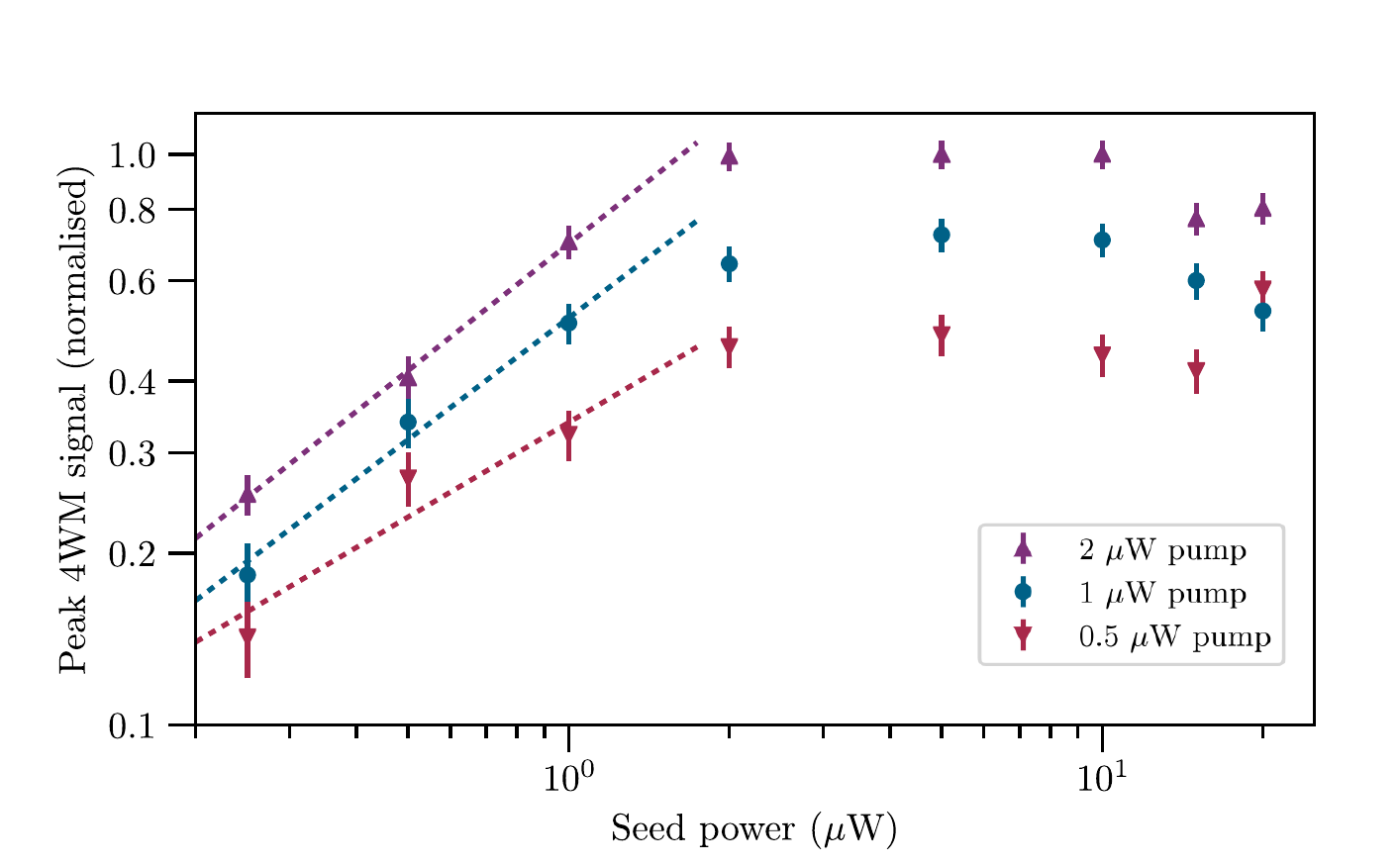}
	\caption{Four-wave mixing signal as a function of the power of the seed beam. The three curves correspond to pump powers of 2~$\mu$W (purple $\blacktriangle$), 1~$\mu$W (blue $\bullet$)  and 0.5~$\mu$W (red $\blacktriangledown$). After an initial sharp rise in signal with seed power, the signal saturates. The peak signal power at saturation is higher for the different pump powers, however the saturation point is unchanged.} 
	\label{4WM-vs-power-seed}
\end{figure}

Fig. \ref{4WM-vs-power-seed} shows the peak signal intensity as a function of seed power, for 3 different pump powers.
Increasing the seed power initially results in a sharp increase in the 4WM signal, but this quickly saturates after a seed power of around \mbox{P$_\text{sat} $= 2.5~$\mu$W}. The initial increase can be fitted with power laws with exponents of 0.74 $\pm$ 0.03, 0.7 $\pm$  0.1 and 0.6 $\pm$ 0.1, for the pump powers of 2~$\mu$W, 1~$\mu$W and 0.5~$\mu$W respectively, as shown by the dotted lines in the figure.
With additional pump power, the peak signal increases, but P$_\text{sat}$ does not change. For very high seed powers, the signal decreases 
which we attribute to power broadening of the pump transition.

\begin{figure}[t]
	\centering
	\includegraphics[width=\columnwidth]{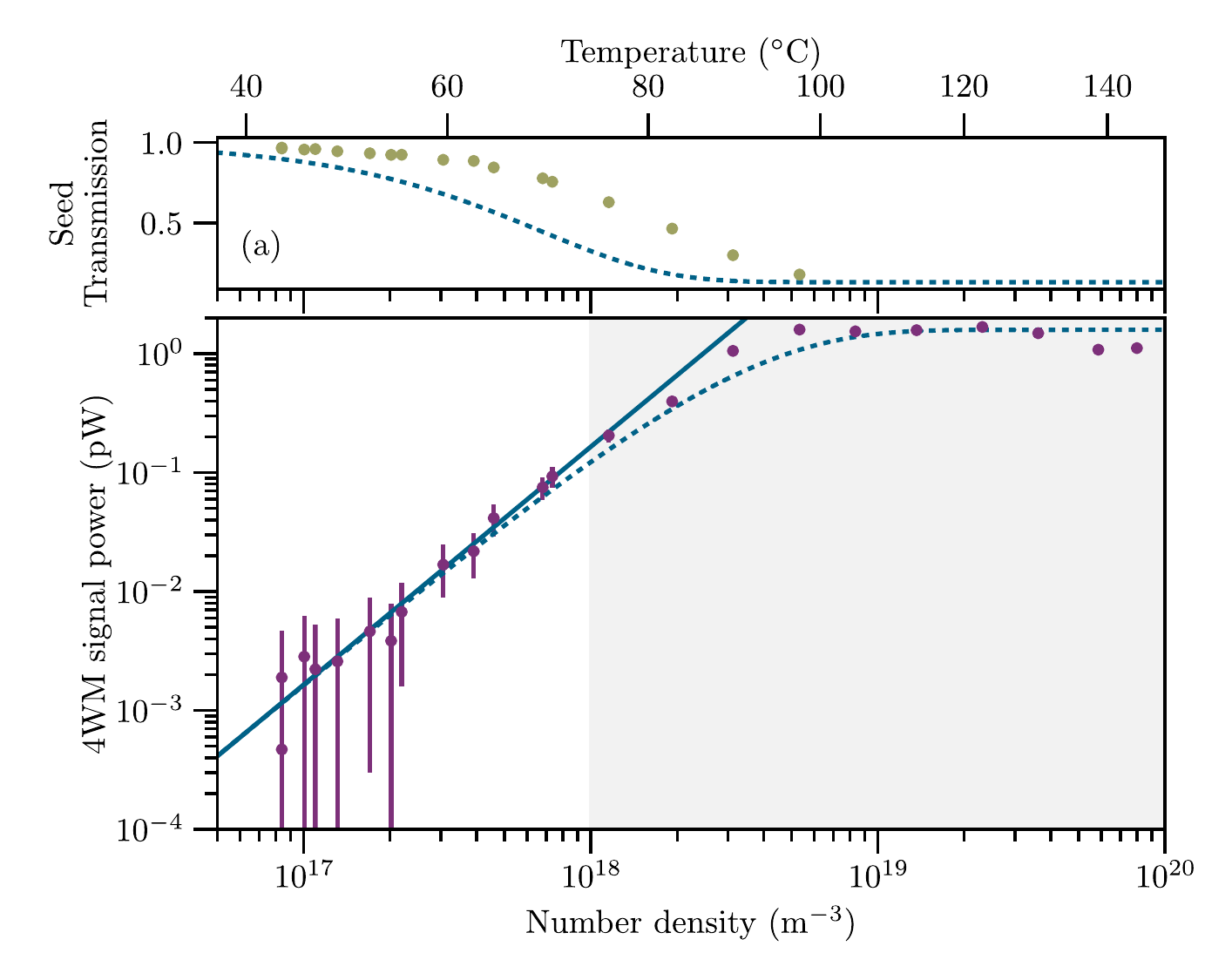}
	\caption{Resonant four-wave mixing peak power and seed transmission as a function of atomic number density and temperature. After an initial quadratic rise (solid blue line), the maximum signal is reached at a temperature of $\sim90^{\circ}$C. The dotted line in (b) is a fit using equation~\ref{intensity_curve} and shows good agreement with the data. Comparing the extracted optical depth from this model (dashed line in (a)) suggests the saturation is caused primarily by the pump laser absorption, rather than the seed absorption (olive points in (a)).} 
	\label{4WM-vs-temp}
\end{figure}

\subsection{Dependence on atomic number density}
In Fig. \ref{4WM-vs-temp} we plot the resonant 4WM peak signal as a function of atomic number density, with the corresponding vapour temperature on the top axis (the axes are the same for both panels). The total emitted electric field scales linearly with the number of atoms since 4WM is a coherent effect. Therefore the signal intensity initially scales quadratically with atomic density (solid blue line in panel (b)). A quadratic fit (i.e. with an exponent of 2) for the data in the unshaded region in panel (b) yields a reduced chi-squared value of 0.1 \cite{Hughes2010}, indicating a good fit. If the exponent is allowed to vary in the fit, we obtain a best-fit exponent of 2.1 $\pm$ 0.3. Outside of this region, at densities above $\sim 5 \times 10^{18}$~m$^{-3}$ (temperature around 90$^\circ$C), the 2~mm long vapour becomes optically thick for the ground state transitions, and this reduces the 4WM signal, since the resonant pump and seed beams on the D$_2$ and D$_1$ transitions are strongly scattered by the medium. The extinction can be described by the Beer-Lambert law. When phase-matching is present, the total electric field from all atoms is simply the constructive superposition of the individual fields. The intensity is then given by \cite{Huber2014}

\begin{equation}
I \propto \Big( \frac{N}{d_{\text{opt}}} \Big)  ^2 \Big[ 1 -\text{exp}\Big( -\frac{d_{\text{opt}}}{2} \Big) \Big]^2,
\label{intensity_curve}
\end{equation}
where the medium has a density $N$ and optical density $d_{\text{opt}}$.

The dotted blue line in panel (b) indicates reasonable agreement with the model described by Eq.(\ref{intensity_curve}). With the same optical depth, we expect transmission of the driving fields given by the dotted blue curve in panel (a), which does not agree with the measured seed transmission. This suggests that it is the pump absorption (and not the seed absorption) which is primarily responsible for the reduction in the 4WM signal. This is not surprising as the optical density is higher for the D$_2$ line (pump) than the D$_1$ line (seed). In the high density regime, the absolute signal power is on the order of 1 pW, compared to the seed power of ~1uW. The low conversion efficiency can be partly attributed to the low seed and pump powers leading to a low non-linear susceptibility and the low detection efficiency of {\raise.17ex\hbox{$\scriptstyle\sim$}}10\%.

\vspace{-1em}
\section{Conclusions}

We have presented an experimental study of seeded non-degenerate four-wave mixing in the hyperfine Paschen-Back regime where the relevant atomic energy levels are reduced to a simple four-level system. The applied magnetic field removes multi-path interference and therefore allows for quantitative agreement with a simple theoretical model based on 4-level optical Bloch equations, even in the regime of strong driving.
We have investigated how the four-wave mixing signal depends on laser detunings, powers and the atomic density of the vapour, finding excellent agreement between theory and experiment.
This study adds to previous related works~\cite{Whiting2015,Whiting2016a} demonstrating how many non-linear optical phenomena can be simplified in thermal vapours by the application of a large magnetic field, allowing for detailed quantitative modelling.
Following on from the development of a range of technologies based on linear atom-light interactions~\cite{Abel2009a,Weller2012b,Zentile2015c,Zentile2015d,Keaveney2016c}, we envision that similar comprehensive modelling of non-linear systems will enable the design and optimisation of future devices. Our approach may enable experiments on strong phase dependence in diamond systems \cite{Kajari-Schroder2007} to be performed using alkali metal vapours.

\section*{Acknowledgements}
We acknowledge financial support from EPSRC (grant EP/L023024/1) and Durham University.
CSA is supported by the EU project H2020-FETPROACT-2014 184 Grant No. 640378 (RYSQ).
The datasets generated during and/or analysed during the current study are available in the Durham University Collections repository (doi:10.15128/r18623hx742).

\bibliographystyle{tfp}
\raggedright
\bibliography{bibliography}

\end{document}